\documentclass[%
 aip,
 amsmath,amssymb,
 reprint,%
]{revtex4-1}

\usepackage{graphicx}% Include figure files
\usepackage{dcolumn}% Align table columns on decimal point
\usepackage{bm}% bold math
\usepackage{import}
\usepackage[utf8]{inputenc}
\usepackage[T1]{fontenc}
\usepackage{mathptmx}
\usepackage{etoolbox}
\usepackage{comment}
\usepackage{xcolor}
\usepackage{enumitem}
\usepackage{booktabs} 
\makeatletter
\def\@email#1#2{%
 \endgroup
 \patchcmd{\titleblock@produce}
  {\frontmatter@RRAPformat}
  {\frontmatter@RRAPformat{\produce@RRAP{*#1\href{mailto:#2}{#2}}}\frontmatter@RRAPformat}
  {}{}
}%
\makeatother
\begin{document}

\preprint{AIP/123-QED}

\title[A versatile beamline for attosecond photoelectron and photon
spectroscopy]{A versatile beamline for attosecond photoelectron and photon spectroscopy}
% Force line breaks with \\
\author{Ronak Narendra Shah}
\affiliation{Institute of Physics, University of Freiburg, Stefan Meier Strasse 19, 79104, Freiburg, Germany.}%
\email{ronak.shah@physik.uni-freiburg.de}
\author{Jahanzeb Muhammad}
\affiliation{Institute of Physics, University of Freiburg, Stefan Meier Strasse 19, 79104, Freiburg, Germany.%\\This line break forced with \textbackslash\textbackslash
}%
\author{Ianina Kosse}
\affiliation{Institute of Physics, University of Freiburg, Stefan Meier Strasse 19, 79104, Freiburg, Germany.%\\This line break forced with \textbackslash\textbackslash
}%

\author{Marvin Schmoll}
\affiliation{Institute of Physics, University of Freiburg, Stefan Meier Strasse 19, 79104, Freiburg, Germany.%\\This line break forced with \textbackslash\textbackslash
}%

\author{Samuel Bengtsson}
\affiliation{Department of Physics, Lund University, PO Box 118, SE-221 00 Lund, Sweden%\\This line break forced with \textbackslash\textbackslash
}%
\author{Riccardo Mori}
\affiliation{Institute of Physics, University of Freiburg, Hermann-Herder-Str. 3, 79104 Freiburg, Germany}
\author{Dominik Hoff}
\affiliation{Single Cycle Instruments, UG(Haftungsbeschränkt) \& Co. KG, Hans-Knoell-Strasse, 07747 Jena, Germany.}
\author{Gerhard G. Paulus}
\affiliation{Institut für Optik und Quantenelektronik, Friedrich-Schiller-Universität Jena, Max-Wien-Platz 1, 07743 Jena, Germany.} 
\author{Fabio Frassetto}
\affiliation{ 
Istituto di Fotonica e Nanotecnologie, CNR, 35131 Padova, Italy.%\\This line break forced with \textbackslash\textbackslash
}%
\author{Luca Poletto}
\affiliation{ 
Istituto di Fotonica e Nanotecnologie, CNR, 35131 Padova, Italy.%\\This line break forced with \textbackslash\textbackslash
}%

\author{Giuseppe Sansone}
    \altaffiliation[Also at ]{Freiburg Institute for Advanced Studies (FRIAS), University of Freiburg, Albertstraße 19, 79104 Freiburg, Germany}
    \affiliation{Institute of Physics, University of Freiburg, Stefan Meier Strasse 19, 79104, Freiburg, Germany.}%

\date{07 November 2026}% It is always \today, today,

\begin{abstract}
We report on the development of a versatile attosecond beamline for ultrafast spectroscopy of atoms and molecules, as well as for the metrology of infrared laser pulses. Using a few-cycle pulse ($\sim$5~fs) for driving the high-order harmonic generation process, we show that the modulation of the harmonic flux in the cut-off spectral region, induced by a weak perturbing field, enables complete electric-field characterization of the perturbing pulse. Phase-sensitive detection with a lock-in amplifier further reveals polarization-dependent effects, observed as modulations at twice the fundamental optical frequency, emphasizing the sub-laser-cycle electronic control inherent to the measurement technique.  In addition, the beamline supports the implementation of two-color (extreme ultraviolet and infrared) phase-sensitive experiments using $\sim$30~fs pulses and a velocity map imaging spectrometer (VMI). This configuration will enable the study of photoemission time delays in atomic and molecular systems. Together, these results establish the beamline as a robust and adaptable platform for probing attosecond dynamics across diverse experimental approaches.
\end{abstract}

\maketitle

\begin{comment}

\begin{quotation}
The ``lead paragraph'' is encapsulated with the \LaTeX\ 
\verb+quotation+ environment and is formatted as a single paragraph before the first section heading. 
(The \verb+quotation+ environment reverts to its usual meaning after the first sectioning command.) 
Note that numbered references are allowed in the lead paragraph.
%
The lead paragraph will only be found in an article being prepared for the journal \textit{Chaos}.
\end{quotation}
\end{comment}
\section{\label{sec:level1}Introduction}

The rise of attosecond science has been enabled by the investigation of laser-matter interaction driven by intense femtosecond laser~\cite{pierreagostiniPhysicsAttosecondLight2004, Krausz_Iavnov_RevModPhys,Corkum2007_review}, together with the development of metrology methods for temporally characterizing these pulses~\cite{orfanosAttosecondPulseMetrology2019a}. This comprises the experimental demonstration of the reconstruction of attosecond beating by the interference of two-photon transitions (RABBIT) approach for trains of attosecond pulses and the attosecond streaking technique for isolated attosecond pulses (IAPs)~\cite{paulObservationTrainAttosecond2001, hentschelAttosecondMetrology2001a, goulielmakisDirectMeasurementLight2004}. These metrology techniques further opened avenues to measure electron dynamics in atoms and molecules~\cite{nisoliNewFrontiersAttosecond2009, sansoneElectronCorrelationReal2012, nisoliAttosecondElectronDynamics2017}.% in the Neon atoms, with electrons from the 2p orbital being slower to leave the atom compared to 2s~\cite{Schultze_2s3p, Isinger2s3p}. 

Over the past decades, various experimental setups have been designed to study such electronic processes. Although several experimental instruments have been specifically developed and optimized for advanced methods—such as attosecond coincidence spectroscopy with long acquisition times~\cite{ahmadiCollinearSetupDelay2020b, ertelUltrastableHighrepetitionrateAttosecond2023}, soft X-ray transient absorption at high photon energies~\cite{zinchenkoApparatusAttosecondTransientabsorption2023b}, and apparatuses for probing molecules in the liquid phase~\cite{jordanPhotoelectronSpectrometerAttosecond2015}—there remains a strong need for the development of versatile attosecond beamlines that provide access to multiple spectroscopic techniques in a single setup.

Attosecond technology has made it possible to temporally resolve the optical oscillations of electromagnetic waves in the visible and near-infrared (IR) spectral regions~\cite{goulielmakisDirectMeasurementLight2004}. Whereas the earliest experimental method for visualizing optical oscillations relied on attosecond streaking spectroscopy using a single isolated attosecond pulse, more recent techniques have exploited the ultrafast temporal resolution of the tunnelling ionization process to temporally sample unknown optical fields~\cite{kimPetahertzOpticalOscilloscope2013, wyattAttosecondSamplingArbitrary2016,carpeggianiVectorialOpticalField2017,choTemporalCharacterizationFemtosecond2019}.

The present paper presents the development of an all-optical setup that relies on the yield of the extreme ultraviolet (XUV) harmonics during the generation process of IAP as an experimental observable to characterize the electric field of an arbitrary laser pulse. Supplemented with phase-sensitive detection with a lock-in amplifier, the technique becomes relatively fast and robust against experimental noise, which is prevalent in the generation and subsequent measurement of harmonics. This is further emphasized by the measurement of weak harmonic modulations at twice the fundamental optical frequency, when the polarization of the test laser field is orthogonal with respect to the driving laser field. This emphasizes the ability of our measurement setup to control the sub-cycle electron quantum paths inherent to the measurement technique.

In order to study the electron dynamics in atoms and molecules using the generated XUV harmonics, either as trains of attosecond pulses or IAP, the present paper also highlights the development of a VMI device capable of measuring photoelectron angular distributions with kinetic energies up to 80 eV. The chosen geometry of the repeller and extractor in the design of VMI produced stray electrons after the process of photoionization, referred to henceforth as secondary electrons. Such secondary electrons are detrimental to attosecond photoelectron spectroscopic measurements. The paper also highlights the successful identification and further removal using a high-voltage switch. 

%Starting from a commercially available carrier envelope phase (CEP) stabilized laser pulses (30 fs, 5 mJ at 1 kHz) based on Titanium:Sapphire (Ti:Sa), the paper further highlights the development of a differentially pumped hollow core fiber (HCF), which spectrally broadens the laser pulses to almost an octave and further compressing it with chirped mirrors produced 5 fs laser pulses. 

%Having intense laser pulses at 30 fs and 5 fs allows us to produce trains of attosecond pulses or IAP. Integrating the XUV laser pulses with VMI further facilitates attosecond spectroscopic measurement of atoms and molecules. Integrating a phase-modulated pump-probe scheme allows for attosecond transient absorption measurement schemes adaptable to trains or IAP~\cite{Wirth_ATAS}. Measurement of the electric field of an arbitrary laser pulse further demonstrates the versatility of the developed attosecond beamline. 

\section{\label{sec:level1}Experimental setup}
\subsection{\label{sec:level2} Femtosecond light source}
The driving light source used in this work is a commercially available carrier-envelope-phase (CEP) stabilized Ti:Sa based oscillator and multi-pass amplification system - FEMTOPOWER$^{\text{TM}}$ V-CEP4 from FEMTOLASERS \textit{Produktions GmbH}, delivering approximately 5 mJ, 30 fs laser pulses at a repetition rate of 1 kHz. The root-mean-square (RMS) CEP stability of these pulses was measured to be less than 300 mrad.

\begin{figure}
%\begin{center}
%\import{Figures/}{Fig1.pdf}
\includegraphics[width=\linewidth]{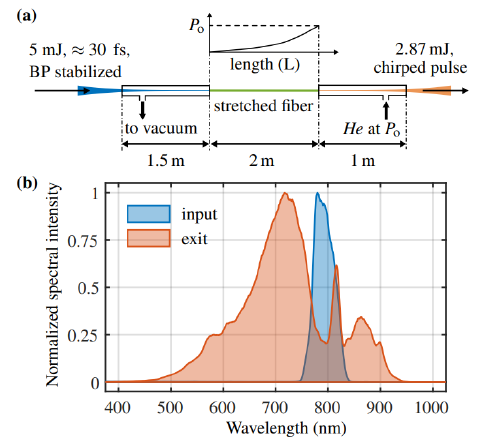}
\caption{Spectral broadening in a HCF. \textbf{(a)} Schematic setup of an in-house developed differentially pumped stretched HCF system. The differential pressure allows for maintaining a gradient of pressure within the fiber that compensates the pulse energy profile along the fiber to reduce the ionization effects. BP: beam pointing. \textbf{(b)} Experimentally measured spectra at the output of the fiber (orange), compared with the fiber input spectra (blue). Adapted from Ref.~\citenum{Ronak_Thesis}} 
\label{Figure1}
%\end{center}
\end{figure}

Based on the time-bandwidth product, achieving the few-cycle femtosecond laser pulses required for the synthesis of an isolated attosecond pulse (IAP) first requires spectral broadening of the laser system output~\cite{HCF_Nagy01012021, HCF_Louisy}. This is accomplished in a differentially pumped hollow-core fiber (HCF) by using the third-order non-linearity of self-phase modulation (SPM). As the name suggests, the fiber has a hollow core, where a non-linear medium, Helium in our case, is pumped at the exit of the fiber and a vacuum environment is maintained at the entrance of the fiber as shown in Fig.~\ref{Figure1} (a). In order to avoid using glass capillaries that are difficult to manufacture over the required length of non-linear medium \textit{i.e.} 2 m in our case, a technology from gas chromatography was borrowed that uses flexible fused silica capillaries covered in polyimide~\cite{HCF_Supplier}. A special system was designed and developed to hold and stretch the fiber, allowing for pumping helium at $P_{\text{o}} = 2$~bar and maintaining a vacuum at the fiber entrance. 

A 2-inch concave mirror having a focal length of 2.5~m was used to couple the input laser pulses having 5~mJ pulse energies. The focal length of the mirror was chosen such that the optimal condition to couple a gaussian beam into the EH$_{11}$ mode of the HCF $w_{\text{o}} \approx 0.65a$, where $w_{\text{o}}$ and $a$ are the beam radius at the laser focus and the HCF core radius, respectively, are satisfied. A beam pointing system (\textit{TEM Messtechnik GmbH}) was implemented before the beam was sent into the fiber to minimize the damage to the fiber entrance due to beam walk-off. A mass flow controller (MKS model $\#$GE50A001103MBV020) was also implemented at the exit of the fiber, which ensured a constant supply of Helium. Owing to this, we were able to obtain stable experimental conditions over the course of the entire day ($\approx 8$~hrs), without any external intervention. The output broadened spectrum is shown in Fig.~\ref{Figure1} (b), and it spans a bandwidth approaching one octave. The laser pulses at the fiber output have pulse energy of 2.87~mJ, indicating an overall efficiency of 57.4$\%$. The output spectrum is blue-shifted; this can be attributed to the complex interplay between SPM, self-steepening, and photoionization-induced broadening~\cite{HCF_Nagy01012021}. 

The output beam from the HCF fiber is collimated using reflective optics and later sent to pulse compression via a chirped mirror compressor having 16 bounces on double-angled chirped mirror - PC1332, Ultrafast innovations GmbH. Each, with a 1.5-inch diameter, has a nominal group-delay dispersion of $-40$ fs$^2$. A higher number of chirped mirrors than required was used to pre-compensate the positive chirp introduced by the transmission optics in the attosecond beamline before generating the XUV harmonics via high-order harmonic generation (HHG). A pair of fused silica wedges having a wedge angle of $4^{\text{o}}$ was used to fine-tune the temporal pulse compression, allowing for optimal pulse compression by compensating the negative chirp introduced in the pulses via the chirped mirrors.

%The verification of the ultrashort laser pulses relied on a stereo-ATI setup based on the phenomenon of above-threshold ionization (ATI). The setup includes two electron detectors placed in opposite directions with respect to the laser focus along the laser polarization axis, where a noble gas, typically Xenon, is ionized [ref - my thesis and Paulus]. The electron yield measured by each detector on a shot-to-shot basis is anti-correlated when the laser pulses are few-cycle in nature. Longer pulses due to their inherent inversion symmetry, where CEP fails to play a role, have no evidence of anti-correlated electron yield [paulus nature 2001]. In practical terms, the method provides an effective means of verifying the few‑cycle nature of the produced laser pulses [ref]. In order to visualize the effect of this anti-correlated electron yield, the photoelectron spectra obtained from the corresponding time-of-flight information of each of the electron detectors are divided into high-energy and low-energy. For each laser shot, a normalized difference is calculated for the high-energy regime for the two photoelectron detectors A$_{\text{HE}}$. Similarly, A$_{\text{LE}}$ is calculated for the low-energy regime. Lissajous patterns arise when A$_{\text{LE}}$ is plotted against A$_{\text{HE}}$ for each laser shot, as shown in Figure \ref{Figure2}.
%

\begin{figure}
%\begin{center}
%\import{Figures/}{Figure2_RSI.pdf_tex}
\includegraphics[width=\linewidth]{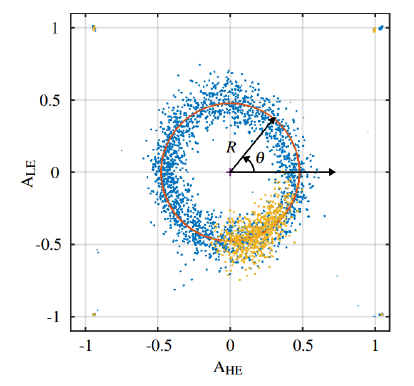}
\caption{Stereo-ATI measurement using few-cycle driving pulses. The blue dots, corresponding to 12520 single laser shots, were obtained by continuously varying the CEP using a 400 Hz saw-tooth waveform supplied to the phase shifter that is used to stabilize the CEP. A circular fit (orange line) results gives the radius of the donut $R=0.48$ units. The yellow dots consisting of 4000 laser shots show the CEP distribution when the CEP is stabilized. Adapted from Ref.~\citenum{Ronak_Thesis}. }
\label{Figure2}
%\end{center}
\end{figure}

The verification of the few-cycle pulse duration relied on a stereo-ATI setup based on the phenomenon of above-threshold ionization (ATI). The setup includes two electron detectors placed in opposite directions with respect to the laser focus along the laser polarization axis, where a noble gas is ionized~\cite{Ronak_Thesis, Paulus2001, wittmannSingleshotCarrierEnvelope2009}. For each laser shot, a normalized difference is calculated for the high-energy regime of the two photoelectron detectors A$_{\text{HE}}$. Similarly, A$_{\text{LE}}$ is calculated for the low-energy regime. Lissajous patterns arise when A$_{\text{LE}}$ is plotted against A$_{\text{HE}}$ for each laser shot, as shown in Fig.~\ref{Figure2}. By proper adjusting the integration intervals for the low and high-energy part of the photoelectron spectrum, the correlated plot resembles an ellipse. In these experiments, few-cycle pulses having pulse energy $\sim 20$ µJ were focused into the vacuum chamber of the Stereo-ATI device via a concave mirror of 25~cm focal length, resulting in intensities $\approx 1 \times 10^{14}$~W/cm$^2$ at the focal spot. Xenon was used as a target gas. The CEP of the single laser pulses is mapped onto the angle $\theta$ in the A$_{\text{LE}}$ vs A$_{\text{HE}}$ plane, as shown in Fig.~\ref{Figure2}. The continuously varying CEP (indicated by blue points), once stabilized (yellow points), enables the measurement of the CEP stability of the generated few-cycle pulses, which is 343~mrad. The radius $R$ of the circle in Fig.~\ref{Figure2} gives a rough indication of the pulse duration of the generated few-cycle laser pulses, estimated in roughly $5.5$~fs~\cite{wittmannSingleshotCarrierEnvelope2009, saylerPreciseRealtimeEverysingleshot2011}.

\subsection{\label{sec:level2} Attosecond beamline}
The attosecond beamline consists of a series of vacuum chambers in which XUV harmonics are generated by the process of HHG, then steered towards the VMI-setup and finally characterized using an XUV spectrometer. 

\begin{figure*}
%\begin{center}
%\import{Figures/}{SectionView_RSI.pdf_tex}
\includegraphics[width=\linewidth]{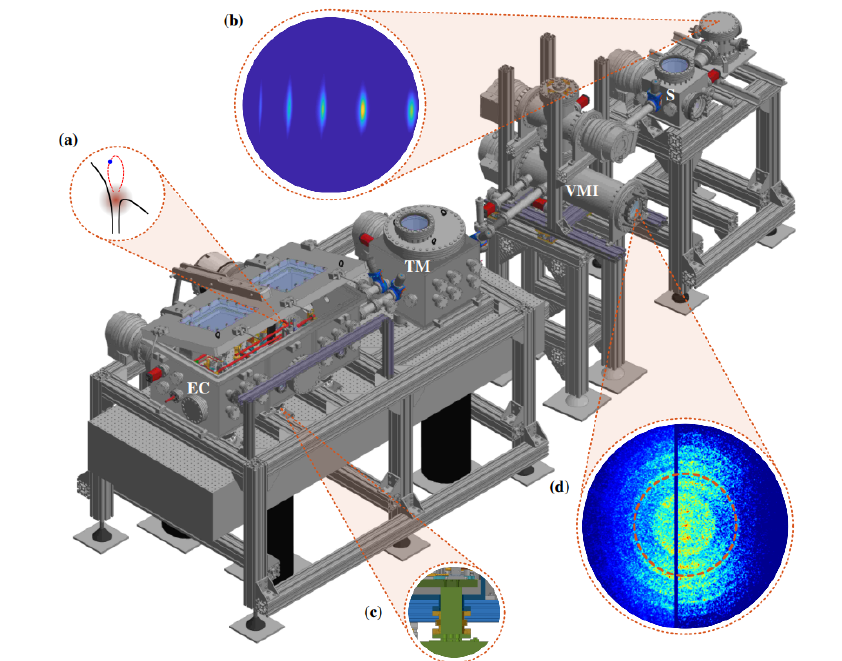}
\caption{Isometric CAD view of the attosecond beamline. \textbf{(a)} Graphical representation of the three-step model, indicating the position where harmonics via HHG are generated. \textbf{(b)} Image at the XUV spectrometer indicating the energy- and angle-resolved typical XUV spectrum. \textbf{(c)} Vibration isolation scheme implemented to decouple the optical setup (optical breadboard and table - green) from the mechanical setup (Bosch profiles and vacuum chamber - blue) via flexible bellows (yellow). \textbf{(d)} Photoelectron spectra obtained by ionizing argon atoms with the generated harmonics in the presence of an IR field at a delay away from zero (left) and near zero delay (right). The orange dotted lines indicate the position of one of the generated harmonics, highlighting the appearance of sidebands between adjacent harmonics. EC: Experimental chamber, TM: Toroidal mirror chamber, VMI: Velocity-map imaging spectrometer, and S: XUV Spectrometer. Adapted from Ref.~\citenum{Ronak_Thesis}.}
\label{Figure3}
%\end{center}
\end{figure*}

A computer-aided design (CAD) view of the beamline developed as a part of this work is shown in Fig.~\ref{Figure3}. The beamline was fabricated at the mechanical workshop of the Physics Institute, University of Freiburg. From left to right with respect to Fig.~\ref{Figure3}, the beamline consists of the following parts:

\begin{itemize}[leftmargin=0.2in]
\item \underline{Experimental chamber (EC)} \\
It houses an optical breadboard of 950 cm $\times$ 430 cm. The large area of the optical breadboard provides versatility to adapt to different experiments as detailed further in sections~\ref{sec:FST} and~\ref{sec:RABIT}. Depending on whether an attosecond pulse train or a spectral continuum (the signature of an IAP) is to be generated, laser pulses of approximately $\sim 30$~fs or $\sim 5$~fs duration, respectively, are coupled into the EC. The laser pulses are focused using a 25 cm focal length concave mirror, which focuses the laser beam close to the tip of the nozzle, which has an opening of 50~µm through which a small fraction of noble gas (Argon in our case) is introduced in the vacuum chamber to efficiently generate and prevent re-absorption of generated harmonics in argon. The nozzle is placed on a three-dimensional motorized stage to optimize the phase matching and selection of short trajectories of the generated harmonics~\cite{Heyl_2017, Gaarde_PM}.

\item \underline{Toroidal mirror chamber (TM)}\\
It houses a toroidal mirror having radii 14330~mm (tangential) and 69.75~mm (sagittal). The entrance and exit arms are 1~m, and the incidence angle is 86$^{\text{o}}$. The mirror is a custom design having a Gold coating procured from Hellma Optik GmbH. Upon entering the TM, the beam passes through a filter wheel equipped with a selection of transmission filters: a 200-nm-thick nickel-mesh-supported aluminum filter, a 100-nm-thick meshless aluminum filter, a 200-nm-thick nickel-mesh-supported tin filter, or an open position that allows the IR beam to pass unfiltered. After the filter wheel and before the toroidal mirror is a holey mirror having a \mbox{3 mm} hole at its center (Bernhard Halle Nachfl. GmbH), which is used to combine XUV that passes through the hole and a delayed IR beam entering from the second entrance of the TM, which are then refocused into VMI by the toroidal mirror.  

\begin{table}[b!]
\caption{\label{tab:table1}Overview of the vacuum levels in the attosecond beamline at standby mode (SB) - where no gaseous load is present, and when in operation (OP) - where a gaseous load is applied. All vacuum pumps are purchased from Pfeiffer Vacuum GmbH\@. VMI-S: The gas source part of VMI, and VMI-D: Dump for the collimated gaseous beam.}
\centering
\begin{tabular}{ccc}
\toprule
\textbf{Chamber} & \textbf{SB [mbar]} & \textbf{OP [mbar]} \\
\midrule
\textbf{EC}    & $8\times 10^{-6}$ & $2\times 10^{-3}$ \\
\textbf{TM}    & $2\times 10^{-8}$ & $6\times 10^{-6}$ \\
\textbf{VMI}   & $3\times 10^{-9}$ & $7\times 10^{-7}$ \\
\textbf{VMI-S} & $7\times 10^{-7}$ & $2\times 10^{-4}$ \\
\textbf{VMI-D} & $3\times 10^{-9}$ & $2\times 10^{-6}$ \\
\textbf{S}     & $1\times 10^{-9}$ & $8\times 10^{-8}$ \\
\bottomrule
\end{tabular}
\end{table}

\item \underline{Velocity-map imaging spectrometer (VMI)}\\
VMI records the momentum distribution of the laser-induced photoionization fragments as a two-dimensional projection, which, due to the cylindrical symmetry of the device, can be reconstructed using the Abel transformation \cite{Eppink1997}. VMI comprises repeller (R), extractor (E), and ground (G) electrodes that, when appropriately biased, image the photoionization fragments onto the MCP–P detector assembly (Hamamatsu F2226-24P-Y012). Additionally, it consists of Einzel lenses L1, L2, and L3, each of which is designed and tested to withstand an electric potential of 10 kV, which provides tunability between the resolution and maximum kinetic energy imaged onto the \mbox{MCP-P.} \mbox{Figure \ref{Figure3} (d)} shows the images captured by a Basler \mbox{Ace 2} camera placed in the air, which records the fluorescence of the phosphor screen. A gaseous mixture of the atoms or molecules under investigation is introduced from above via a nozzle having an opening of 10 µm and collimated using a skimmer with a 400 µm aperture. At the bottom of the setup, a vacuum pump collects the gas jet, enabling reliable operation under high electrical voltages and preventing arcing.

\item \underline{XUV Spectrometer (S)}\\ 
The spectrometer chamber consists of a fixed part and a movable part. The fixed part comprises a cylindrical diverging mirror (gold-plated, with a sagittal radius of $-80$~mm, entrance arm 930~mm, and incidence angle $85^{\text{o}}$) and an aberration-corrected concave grating for a flat-field spectrograph (Hitachi High-Tech Corporation, part\# 001-0640). The movable part houses an \mbox{MCP-P} (Hamamatsu, F2225-11P). A CMOS camera (Hamamatsu, C11440-36U ORCA-spark digital) placed in the air captures the image of a spatially separated series of odd harmonics visible on the phosphor screen of \mbox{MCP-P}, as shown in Fig.~\ref{Figure3}(b). The zero-order diffraction from the grating is physically blocked inside the spectrometer chamber. 
\end{itemize}

Notice that the optical breadboards inside the EC and the TM chambers are directly connected to the optical breadboards through flexible vacuum bellows (see Fig.~\ref{Figure3} (c)),  while the chambers themselves are sustained by a set of profiles sitting on the floor. This solution allows us to decouple sources of vibrations \textit{i.e.}, vacuum pumps, from the optical setup, thus improving the stability of the experimental conditions. 

Each of the vacuum chambers is pumped in two stages by a pre-vacuum pump (multistage root pump) followed by a turbo-molecular pump to achieve the desired ultra-high vacuum (UHV) range. Table \ref{tab:table1} summarizes the final vacuum levels reached in operation and in standby for the individual chambers. 

\section{\label{sec:FST} Field sampling techniques}

The advancement of attosecond metrological tools also facilitated the development of electric field sampling techniques. A distinct advantage of these measurement techniques is that it provides a direct electric field measurement without any reconstruction or deconvolution algorithm. Moreover, it provides access to the complete electric field profile, including the CEP of the laser pulse.

Cross-correlating an IR laser pulse with an IAP enables measurement of the IR vector potential, which serves as the basis for attosecond streaking~\cite{goulielmakisDirectMeasurementLight2004}. Other all-optical techniques rely on the sub-femtosecond window inherent to the process of HHG. In these measurements, a small perturbation of the test field to be measured changes the properties of the XUV radiation produced. Thus, measuring the properties of XUV radiations provides access to the electric field profile of the perturbing test pulse. Several experimental techniques have been developed for the complete temporal characterization of femtosecond laser pulses: in the petahertz oscilloscope, which combines an IR laser pulse with a strong laser pulse producing IAP in an angular geometry, the angular modulation of harmonic radiation depends on the instantaneous value of the IR field to be measured~\cite{kimPetahertzOpticalOscilloscope2013}. In the attosecond resolved interferometric electric-field sampling (ARIES) approach, the observable is the modulation of the HHG cut-off, which depends on the relative delay between the collinearly propagating driving and unknown pulses~\cite{wyattAttosecondSamplingArbitrary2016}. Finally, the XUV interferometric approach is based on the generation of two IAP pulses; when the IR field to be measured is overlapped with one of the IAP sources produces amplitude and phase modulations of the harmonic fringes in the far-field, which reveals the value of the unknown electric field~\cite{carpeggianiAttosecondElectronicRecollision2018, carpeggianiVectorialOpticalField2017}. In contrast to these approaches, the Tunneling ionization with a perturbation for the time-domain observation of an electric field (TIPTOE) approaches measures the electrons that miss the parent atom or the ionization rate to reconstruct the unknown pulse~\cite{choTemporalCharacterizationFemtosecond2019}.

The electric field characterization techniques described above are not an exhaustive list, and a comparison between the different techniques is beyond the scope of this work. Within this section, an in-house developed all-optical field sampling technique is discussed. This technique combines the unique benefits of ARIES and XUV interferometry, offering both the simplicity of ARIES’ experimental setup and the amplitude modulations in the cut-off region of the harmonic spectral continuum, which is a characteristic of XUV interferometry. Such a selection allows us to successfully implement a phase-sensitive lock-in detection to improve the signal-to-noise ratio (SNR) and, at the same time, have high scan rates~\cite{Ronak_Thesis}.

\subsection{\label{sec:level2} Measurement principle}

\begin{figure}[t!]
%\begin{center}
%\import{Figures/}{XUV_I.pdf_tex}
\includegraphics[width=\linewidth]{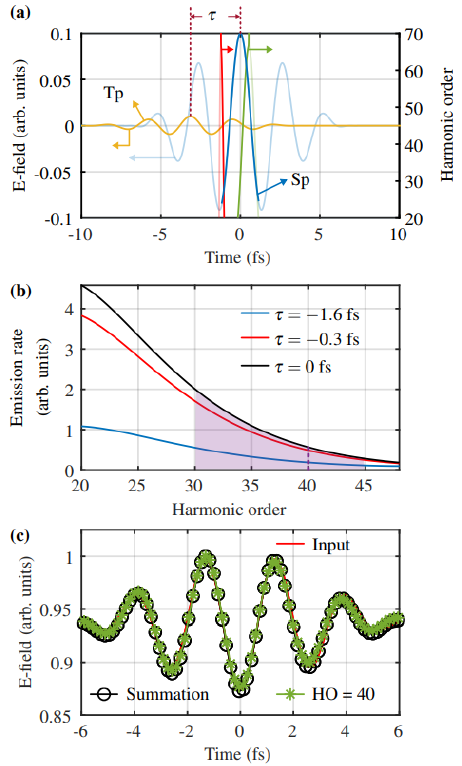}
\caption{SFA model for the developed field sampling technique. \textbf{(a)} Electric field profile of the sampling pulse (Sp, blue) and the test pulse (Tp, yellow) at a delay $\tau$ with respect to the Sp. Ionization (red) and recombination (red) curves corresponding to the two classes of trajectories (short and long) leading to the highest photon cut-off are presented. Only the short trajectory (thick lines) is selected in the non-adiabatic model of HHG to mimic the short-path selection occurring in the gas jet. \textbf{(b)} Harmonic emission rate for different $\tau$. \textbf{(c)} Selecting the single harmonic order (HO) = 40  (black open circles) or summing the HO from 30 to 45  (green asterisk), overlay well with the Input Tp (red line). An amplitude scaling was performed to overlap the traces to have the same amplitude limits of the electric field. A pulse duration of 5 fs pulse at the central wavelength of \mbox{$\lambda_{\text{o}} = 800$ nm} is used in the simulations. Note that the Time axis in (c) represents the delay $\tau$, whereas the Time axis in (a) is absolute time. Adapted from Ref.~\citenum{Ronak_Thesis}.}
\label{Figure4}
%\end{center}
\end{figure}

Within our technique, the laser pulse that provides the ultrashort temporal gate is indicated as the sampling pulse (Sp), and the laser pulse that needs to be characterized is called the test pulse (Tp), as shown in Fig.~\ref{Figure4} (a). The net electric field $\textbf{E}_{\text{net}}$ that participates in the HHG process can now be written as:

\begin{equation}
    \textbf{E}_{\text{net}}(t, \tau) = \textbf{E}_{\text{Sp}}(t)  + \textbf{E}_{\text{Tp}}(t, \tau),
\end{equation}

where $\tau$ is the delay between the Sp and the Tp. Within the framework of strong-field approximations (SFA) in the non-adiabatic limit of a few-cycle laser pulses, the condition of an ultrashort temporal gate created by the Sp is imposed by selecting only a single short quantum trajectory corresponding to the one having the highest cut-off. In this case, the dipole spectrum $x(\omega)$ of the single-atom response can be written as \cite{PRA-Lewenstein-1994,sansoneNonadiabaticQuantumPath2004,sansoneControlLongElectron2006,sansoneQuantumPathAnalysis2009}:
\begin{widetext}
\begin{equation} \label{SFA_Equation_Exp}
\begin{split}
     x(\omega) = 
     \frac{i\, 2\pi}{\sqrt{\text{det}(S'')}}\Bigg[ \frac{\pi}{\epsilon + i\frac{(t_{s} - t{s}')}{2}} \Bigg]^{\frac{3}{2}}\underbrace{\textbf{E}_{\text{net}}(t_{s}')\textbf{d}(\textbf{p}_{s} - \textbf{A}_{\text{net}}(t_{s}'))}_{\text{Ionization}} \underbrace{e^{[i\Theta(\textbf{p}_{s}, t_{s}, t_{s}')]}}_{\text{Continuum}}\underbrace{\textbf{d}^{*}(\textbf{p}_{s} - \textbf{A}_{\text{net}}(t_{s}))}_{\text{Recombination}},
\end{split}
\end{equation}
\end{widetext}

\noindent where $\epsilon$ and $\textbf{d}$ are a regularization constant and the Gaussian form of the dipole matrix element, respectively. The phase term $\Theta(\textbf{p}, t, t')$ is given by:

\begin{equation}\label{SFA_Theta_Equation}
\begin{split}
    \Theta(\textbf{p}, t, t') &= \omega t - S(\textbf{p}, t, t')\\
    &= \omega t - \int_{t'}^{t}dt''\Bigg(\frac{[\textbf{p} - \textbf{A}(t'')]^{2}}{2} + I_{p} \Bigg).
\end{split}
\end{equation}

In the above equation, the canonical momentum \mbox{$\textbf{p} = \textbf{v} + \textbf{A}(t)$} was used, where $\textbf{A}(t)$ is the vector potential of the applied electric field, and $I_{p}$ is the ionization potential of the target atom. $(\textbf{p}_{s}, t_{s}, t_{s}')$ are the saddle-point solutions associated with the complex electron quantum paths that mostly contribute to the harmonic spectrum. $S$ is the dipole phase and det($S''$) is the determinant of the matrix of the second time derivative of the dipole phase given by:

\begin{equation}   
     \text{det}(S'') = \Bigg [ \frac{\partial^{2}S}{\partial t \partial t'} \Bigg]^{2} - \frac{\partial^{2}S}{\partial t^{2}}\frac{\partial^{2}S}{\partial t'^{2}}.
\end{equation}

$x(\omega)$ allows us to calculate the harmonic emission rate $\propto |x(\omega)|^{2}$ for different delay $\tau$ as shown in Fig.~\ref{Figure4} (b). Changing the delay between the Sp and the Tp, a variation in the harmonic emission rate is seen. The variations can be understood as a manifestation of the Tp, which perturbs the tunneling ionization rate, the complex electron trajectory in the continuum, and the recombination probability in the HHG process of the Sp, as seen from Equation \ref{SFA_Equation_Exp}. To quantify these variations, we perform a summation from harmonic order (HO) 30 to 45, as shown by the shaded region in Fig.~\ref{Figure4} (b), or we can track the variation of a single harmonic at HO = 40. When overlaid with the applied Tp, a relatively good agreement between the applied Tp and the harmonic emission rate obtained via solving the non-adiabatic model of HHG is achieved, as shown in Fig.~\ref{Figure4} (c). A linear variation of the harmonic emission rate as a function of the applied perturbation in terms of Tp thus allows us to directly measure the electric field profile of any arbitrary applied test field, within the limit of this linear regime.

Apart from Tp being a perturbation, if the Tp exhibits spectral components lower than 200~nm, the optical period of such oscillating components will start being comparable with the ultrashort temporal gate offered by tunnel ionization in the HHG process, thereby leading to undersampling of the unknown field~\cite{carpeggianiAttosecondElectronicRecollision2018, kimPetahertzOpticalOscilloscope2013}. In the simulation results presented in Fig.~\ref{Figure4}, the laser intensity of the Sp and the Tp in the harmonic generation process was $3.5\times 10^{14}$ W/cm$^{2}$ and $3.5\times 10^{10}$ W/cm$^{2}$, respectively. The high sensitivity of the harmonic emission rate to even small changes of the electric field of the Tp becomes a significant advantage of this measurement technique, enabling the detection of the Tp with exceptionally low pulse energy - just a few nanojoules. In the present simulations, the polarization of the
Tp is oriented in the direction of the polarization of the Sp. The effects of increasing amplitude and changing the polarization of the Tp with respect to the Sp are discussed later in Section~\ref{sec:2omega}.

\subsection{\label{sec:level2} Experimental implementation}

\begin{figure*}
%\begin{center}
%\import{Figures/}{Figure5_n.pdf_tex}
\includegraphics[width=\linewidth]{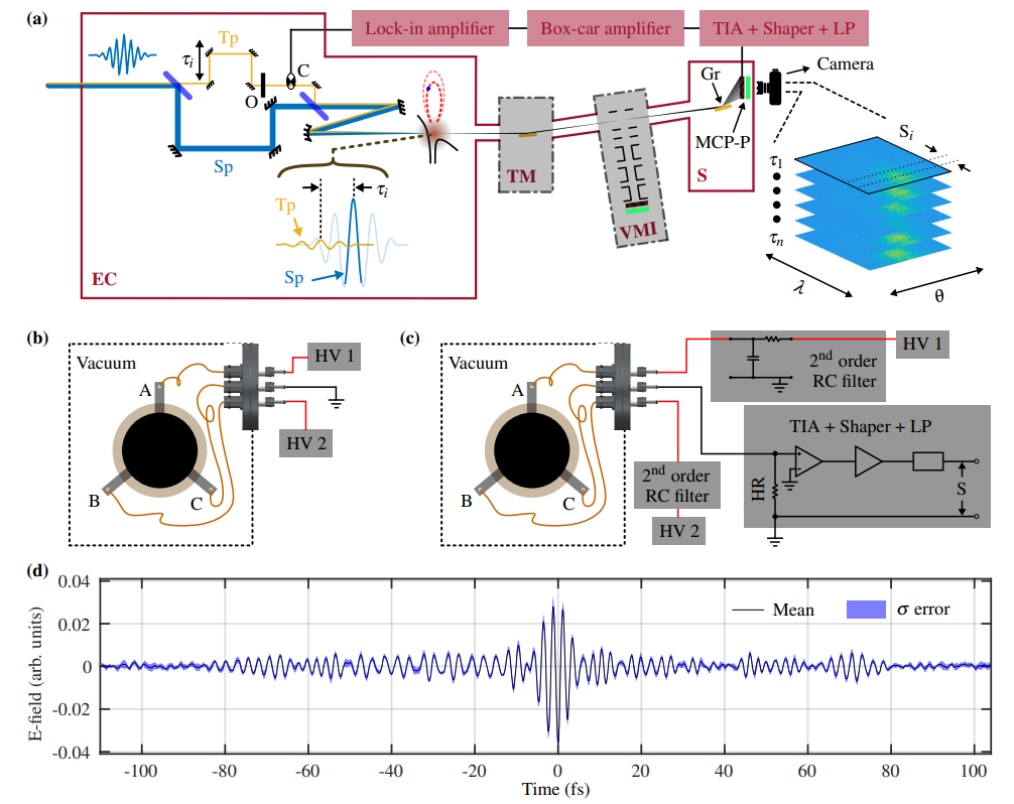}
\caption{Experimental implementation of the field sampling technique. \textbf{(a)} Schematic of the setup, including a Mach–Zehnder interferometer integrated within the EC and the detector S used to measure the harmonic signal. O: Optic placed $(\lambda/2)$ for rotating polarization; C: chopper; TM: toroidal mirror; S: XUV spectrometer; Gr: grating. \textbf{(b)} Typical electrical connections to MCP-P. \textbf{(c)} Modified connections to the MCP-P, where the electrical variations are measured via a transimpedance amplifier (TIA) and the high-frequency noises are further suppressed with a low-pass filter (LP). \textbf{(d)} Long-term measurements showing the mean and $1\sigma$ confidence interval calculated from eight measurements spanning approximately six hours. Adapted from Ref.~\citenum{Ronak_Thesis}.}
\label{Figure5}
%\end{center}
\end{figure*}

As discussed in the preceding section, accurately characterizing the electric field of a few‑cycle laser pulse requires a few key elements: (a) a high‑intensity driving pulse capable of producing an XUV spectral continuum, (b) an interferometric setup that can precisely control the delays between the Sp and Tp, and (c) an appropriate measurement platform to measure the generated harmonics. Intense few-cycle pulses, a Mach-Zehnder interferometer incorporated within the EC, together with the XUV spectrometer, satisfy these criteria, as shown in Fig.~\ref{Figure5} (a), and thus allow us to use the developed attosecond beamline for the temporal reconstruction of unknown electric fields. 

The Mach-Zehnder interferometer is constructed with the help of broadband beam splitters to preserve the few-cycle pulse duration of both pulses. The ratio of the pulse energy of Tp with respect to that of Sp is $\le 0.01$, thereby ensuring that Tp is a perturbation in the HHG process. Individual motorized IRIS in each arm further allow to tune the pulse energy ratio. The delay between the Tp and the Sp is adjusted using a piezo actuator (6PX100CAP, piezosystem jena GmbH). The XUV spectral continuum is typically generated via the ionization-gating technique, in which the phase matching in the HHG process is tuned to limit harmonic generation to one optical cycle of the Sp~\cite{ABEL20099_ionizationGating}. A CMOS camera placed in the air allows us to observe a 2D XUV spectrum with respect to its wavelength $\lambda$ and divergence angle $\theta$. As shown in \mbox{Fig.~\ref{Figure5} (a)}, a series of these images for each relative delay $\tau_{i}$ allows summing the spectral amplitude in the cut-off region $\text{S}_{i}$ of the XUV spectrum, thereby characterizing the Tp. 

The strong non-linear nature of the HHG process, combined with the avalanche-based detection of XUV photons in the MCP-P, makes the measurement particularly susceptible to noise. As a result, longer camera integration times are required at each delay position to reduce fluctuations through averaging. Although this approach is workable, acquiring a full dataset of 2000 points with a camera-based system would take approximately 33 minutes, rendering field characterization relatively time-consuming. A more efficient strategy to reduce the detrimental effect of experimental noise involves the use of phase-sensitive lock-in detection. By incorporating an optical chopper and synchronizing detection to its reference frequency and phase, only the desired signal component is extracted. Noise contributions at other frequencies are effectively filtered out, leading to a significant improvement in the signal-to-noise ratio.

An effective implementation of lock-in detection requires acquiring the XUV spectral signal at the laser repetition rate (1 kHz in this case). Although the CMOS camera used in the setup supports line readout speeds up to 2008.9 frames per second, the phosphor screen (Hamamatsu P43), with a $10\%$ decay time of approximately 1 ms, can introduce aliasing effects that limit accurate measurements. To overcome this limitation, a method based on monitoring the current in the MCP power lines was developed. Figure~\ref{Figure5} (b) shows typical connections made to MCP-P assembly.  MCP-P works on avalanche detection, which is used to detect single individual particles, XUV photons in our case. As the XUV photons impinge on the channels of the MCP, they produce an electron avalanche, which is provided by the MCP biasing current. Thus, a variation in the XUV photons would result in minute variations in the current flowing through the power lines of the MCP. A developed read-out device, Transimpedance amplifier (TIA), is connected to the ground line of the MCP-P assembly, as shown in Figure \ref{Figure5} (c). A $2^{\text{nd}}$ order RC (resistor-capacitor) filter is connected on the high-voltage lines to limit electrical noise. The TIA detects variations in the electrical current which, when synchronized with the laser repetition rate, directly correspond to the number of incident photons.

The resulting signal from TIA, which can be processed at frequencies up to 500 kHz, is well-suited for lock-in detection. A schematic representation of this avalanche-based detection scheme is shown in Fig.~\ref{Figure5} (c). The developed read‑out electronics was followed by a boxcar amplifier (SR200, Boxcar Averager System, Stanford Research Systems) to further suppress noise fluctuations before lock‑in detection (SR810, DSP lock-in amplifier, Stanford Research Systems), which was phase referenced to an optical chopper (SR540, Chopper controller, Stanford Research Systems), set at 400 Hz. A typical scan of 2000 data points can now be completed in approximately 6.6 minutes. This corresponds to a 4.5‑fold increase in acquisition speed relative to the camera‑based method. 

Using the developed system, the statistical mean and $1\sigma$ confidence interval for eight consecutive measurements of few-cycle Tp are presented in Fig.~\mbox{\ref{Figure5} (d)}. In these measurements, negative time indicates pre-pulse. Using the electric field profile, the pulse duration can now be determined to be \mbox{5.58~fs} (FWHM). The measurements also reveal that the higher-order phase terms ($\ge$ third order) are not properly compensated, manifesting as satellite pulses in the pre-pulse as well as the post-pulses. 

Each of the measurements used in the statistical analysis in Fig.~\ref{Figure5} (d) was $\approx45$ minutes apart. However, owing to the lack of beam pointing stabilization at the entrance of the vacuum chamber, a one-directional drift was observed in these measurements, due to a positional drift occurring over the course of these measurements \textit{i.e.} six hours, which causes the electrical field profiles to slip the zero delay over time. The amount of this delay slip is characterized by an error signal, which is calculated from the difference in amplitudes between the two measured experimental traces. A delay shift program is employed to find and correct for the temporal shift $\delta$t which minimizes the error signal given by:

\begin{equation}\label{Eq:Error_Signal}
\text{Error signal} = \sum_{i}\bigg| \big|\text{E}_{1}(t_{i})\big| - \big|\text{E}_{2}(t_{i} + \delta t)\big| \bigg |,
\end{equation}

\noindent where E$_1$ and E$_2$ are individual electric field measurements, and a summation over each delay step $i$ is performed to get the error signal. Considering the first and last measurements were $\approx6$ hours apart, a maximum shift $\delta$t of 3.5 fs was observed. Based on the amount of shift, the edges of the measurements were truncated to minimize measurement error. Given the tight $1\sigma$ confidence interval around the mean shown in \mbox{Fig.~\ref{Figure5} (d)}, it can be concluded that the developed measurement system is robust and that the procedure for the extraction of the delay shift is robust and efficient. Furthermore, the collection of measurements indicates the overall spectral stability of the generated few-cycle laser pulses.

\subsection{\label{sec:2omega} Sub-cycle electron trajectory control}

In the developed measurement technique, Tp acts as a perturbation in the harmonic generation process of the Sp. These perturbations manifested themselves as modulations in the harmonic flux, which were used to characterize the Tp itself. In other words, the Tp was able to modulate the quantum pathways of the electrons involved in the HHG process of Sp. In the present section, this control is further highlighted in terms of the polarization dependence of Tp with respect to the Sp.

\begin{figure}[b!]
%\begin{center}
%\import{Figures/}{Figure6.pdf_tex}
\includegraphics[width=\linewidth]{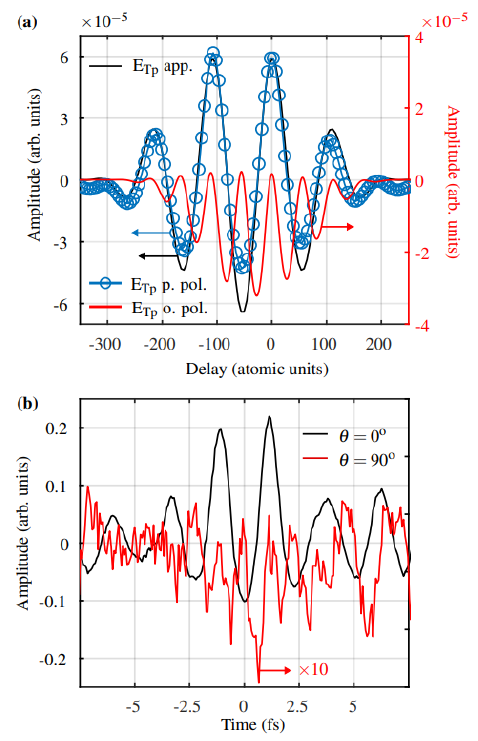}
\caption{Polarization dependence in field sampling \textbf{(a)} SFA-based theoretical model where Tp and Sp were each 5-fs pulses at the central wavelength $\lambda_{\text{o}} = 800$ nm. The black curve indicates the electric field of the Tp introduced in the simulations, and the two special cases retrieved field oscillations for parallel (blue curve) and orthogonal polarizations (red curve) of Tp with respect to Sp. The electric field amplitude of Tp was 0.01 atomic units in the respective polarization state, and the electric field amplitude of Sp was \mbox{0.1 atomic units.} \textbf{(b)}  Experimental results for parallel polarization state: $\theta = 0^{\text{o}}$ (black curve), and orthogonal polarization state: $\theta = 90^{\text{o}}$ (red curve). An error of $\pm 2^{\text{o}}$ is expected in experimental results in $\theta$ due to uncertainty in the rotation stage used. Adapted from Ref.~\citenum{Ronak_Thesis}.}
\label{Figure6}
%\end{center}
\end{figure}

The SFA model developed in the previous sections provides a good theoretical framework for understanding the effects of polarization of Tp. Towards this end, the integral $x(\omega)$ given in Equation \ref{SFA_Equation_Exp} needs to be solved, with the total electric field \textbf{E}$_{\text{net}}(t, \tau)$ now given by:

\begin{equation}
\begin{split}
    \textbf{E}_{\text{net}}(t, \tau) &= \textbf{E}_{\text{Sp}}(t) +  \textbf{E}_{\text{Tp}}(t, \tau)   \hspace{5mm} \text{or},\\ \vspace{10mm}
     \begin{bmatrix}\text{E}_{\text{net}_{x}}(t, \tau)\\ \text{E}_{\text{net}_{y}}(t, \tau)\end{bmatrix}  & = \begin{bmatrix}\text{E}_{\text{Sp}_{x}}(t)\\ \text{E}_{\text{Sp}_{y}}(t)\end{bmatrix} + \begin{bmatrix}\text{E}_{\text{Tp}_{x}}(t, \tau)\\ \text{E}_{\text{Tp}_{y}}(t, \tau)\end{bmatrix}.
    \end{split}
\end{equation}

In further discussions, Sp is considered to be linearly polarized along the x-direction,
thereby: E$_{\text{Sp}_{\text{y}}}(t) = 0$. Varying the delay $\tau$ between the Tp and Sp, Fig.~\ref{Figure6} (a) shows the effect of the polarization of Tp on the harmonic emission rate, when Tp is either polarized along the Sp - blue open circles or orthogonal to Sp - red solid line. From the figure, two inferences can be drawn:
\begin{itemize}[leftmargin=0.15in]
\item[(i)] When Tp is polarized along Sp, the retrieved field via SFA (blue open circles) is asymmetric in amplitude. Compared to Fig.~\ref{Figure4} (c), where the ratio of amplitude of Tp/Sp $= 0.01$, the ratio of amplitude in Fig.~\ref{Figure6} (a) is Tp/Sp $= 0.1$, thereby making the Tp no longer a perturbation in the harmonic generation process of Sp. Thus, in the field sampling measurements envisaged to measure the electric field of Tp, the ratio of pulse energy of Tp/Sp must be set to $\le0.01$. 
\item[(ii)] When Tp is polarized orthogonal to Sp, the modulation of the cut-off harmonic yield simulated via SFA (red curves) exhibits an oscillation at twice the expected period and an amplitude asymmetry with predominantly negative oscillations, henceforth called as $2\omega$ oscillations. Although not shown here, visualizing the $2\omega$ oscillations requires that the Tp must be perfectly orthogonal in its polarization state with respect to the Sp. Even a small mismatch in polarization (less than $2^{\circ}$ degrees) leads to the measurement of field oscillations corresponding to the component of the Tp electric field projected along the polarization direction of Sp. This property makes this technique a tool to measure the complete time-dependent polarization state of Tp. Such measurements of the complete polarization state of an arbitrary laser pulse were shown previously using XUV interferometry~\cite{carpeggianiVectorialOpticalField2017}.
\end{itemize}

A more intuitive understanding of the $2\omega$ oscillations can be obtained by considering the harmonic generation process within the framework of the three-step model \cite{Corkum_threestepmodel_1994}. For parallel polarization, the overall field remains linearly polarized. The Tp introduces a linear perturbation to the HHG process within a certain intensity regime. In contrast, when Tp is polarized orthogonally to Sp, its presence modifies the ellipticity of the combined driving field involved in HHG. Under elliptical polarization, the returning electron no longer revisits the parent ion precisely, but instead passes nearby during its trajectory. This reduces the probability of recombination and consequently lowers the harmonic yield. As a result, a harmonic signal that would otherwise remain constant in the absence of Tp becomes suppressed when an orthogonally polarized Tp is applied, due to the decreased likelihood of electron recombination. This leads to an overall reduction in the signal, with predominantly negative contributions in the modulation. 

Moreover, with orthogonal polarization, the electron wave packet is deflected away from the parent ion during both the positive and negative half-cycles of Tp. Consequently, the modulation of the harmonic yield occurs twice within one optical cycle of Tp, leading to $2\omega$ oscillations. Increasing the field strength of Tp enhances this effect, resulting in larger amplitudes of the observed $2\omega$ oscillations.

The high signal-to-noise ratio achieved through the implementation of phase-sensitive lock-in detection makes the developed technique particularly well suited for experimentally resolving the $2\omega$ oscillations. The polarization of the Tp was rotated with a half-wave plate (Broadband $\lambda/2$ wave plate, AR-coated, $500 - 900$~nm, Bernhard Halle Nachfl. GmbH.), which is placed in the Tp arm of the Mach-Zehnder interferometer - at optic (O) in Fig.~\ref{Figure5} (a). Figure~\ref{Figure6} (b) shows the experimentally obtained field oscillations for the parallel and orthogonal polarization states of Tp with respect to Sp. At $\theta = 90^{\text{o}}$, the measured oscillation shows a clear $2\omega$ oscillation, thereby indicating the successful ability to control and resolve the effects of the two-dimensional electron trajectories with the developed measurement technique.

\begin{figure*}
%\begin{center}
%\import{Figures/}{All_In_One.pdf_tex}
\includegraphics[width=\linewidth]{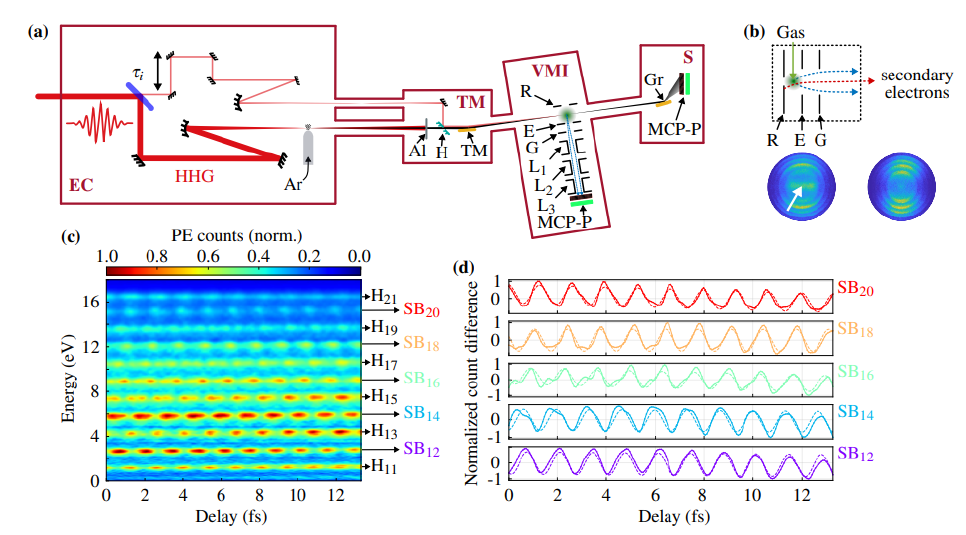}
\caption{Experimental implementation of pump-probe experiments with attosecond temporal resolution using a VMI detector. \textbf{(a)} Schematic of the setup, including a Mach–Zehnder interferometer integrated within the EC and TM chambers, and the VMI and XUV spectrometers. Al: aluminum filter; H: holey mirror; TM: toroidal mirror; R: repeller; E: extractor; G: Ground; Gr: XUV grating; L1, L2, L3: electrode lenses; S: XUV spectrometer. \textbf{(b)} Effect on the center of the VMI images of secondary electrons (red dashed lines) generated by ion impact on the repeller electrode. The primary electrons generated in the interaction region are shown by blue dashed lines. \textbf{(c)} Two-dimensional photoelectron spectra acquired by changing the relative delay between the XUV and IR field in steps of $67$ as. \textbf{(d)} Oscillations of the sideband signals (solid lines) and corresponding fits (dashed lines) using a cosine function oscillating at 2$\omega$.}
\label{Figure7}
%\end{center}
\end{figure*}

\section{\label{sec:RABIT} Photoelectron spectroscopy of two-color photoionization}
The attosecond beamline also enables pump–probe experiments based on photoelectron spectroscopy with attosecond temporal resolution. This capability is illustrated by measuring the oscillations of sideband signals produced in a two-colour photoionization process, in which an XUV harmonic comb is combined with a synchronized IR pulse~\cite{paulObservationTrainAttosecond2001}. The pulses were split in the HHG chamber using an $80$R$:20$T beam splitter as shown in Fig.~\ref{Figure7}a. In one arm of the interferometer, odd harmonics of the fundamental radiation were generated in argon, while the $20\%$ transmitted fraction of the IR field was delayed using a piezoelectric stage with a resolution of about $67$ as. This delay stage was used to adjust the relative delay in the two-colour photoionization process. The IR co-propagating with the XUV radiation was filtered out using a 100-nm-thick Al filter. The XUV radiation and the delayed IR field were recombined using a holey mirror placed at 45$^{\circ}$. A toroidal mirror then refocused both pulses in the interaction region of the VMI spectrometer by forming an image of the XUV source point with unitary magnification. In order to match the divergence of the XUV radiation, the delayed IR field was also focused in a position exhibiting the same distance from the toroidal mirror as the interaction region for the HHG process. The intermediate focus of the IR field also allows for placing additional irises in the beam path to improve the differential pumping between the HHG chamber and the toroidal mirror chamber.

The VMI spectrometer consisted of a repeller, extractor, and ground electrode. The additional \textit{Einzel} lenses L1, L2, and L3 were not used during the measurements presented below and were all grounded. In the absence of any IR field in the VMI, it was noticed that apart from the typical harmonic rings that form our primary electrons, a blob of low-energy electrons was observed as shown in Fig.~\ref{Figure7}(b). The low-energy secondary electrons were a result from repeller electrode due to the impact of high-kinetic-energy ions accelerated by the field present between the repeller and extractor. Because secondary electrons originate from the repeller electrode, they are expected to arrive later than the primary electrons. To suppress their contribution, a pulsed voltage was applied to the microchannel plate (MCP) (see Fig.~\ref{Figure7}b), preventing amplification of the delayed secondary electrons.

The XUV spectrometer placed downstream was used for online monitoring of the characteristics of the XUV radiation during the acquisition of the photoelectron spectra.

The Abel-inverted two-dimensional photoelectron trace clearly reveals periodic oscillations of the main lines, corresponding to the photoelectron released by the absorption of one XUV photon and of the sidebands, due to the combined action of the XUV and IR field (see Fig.~\ref{Figure7}c). These oscillations are characterized by a strong oscillation at frequency 2$\omega$. However, the oscillations of the sidebands presented in Fig.~\ref{Figure7}d exhibits also clear deviations from simple cosine oscillations, as indicated by the corresponding cosine fits of the experimental signals (solid continuous lines). These variations are attributed to the exchange of several photons with the IR field that contribute additional modulation frequencies at 4$\omega$ and 6$\omega$ to the sideband signal \cite{swobodaIntensityDependenceLaserassisted2009}. This is consistent with the estimated high IR intensities of $8.1\times10^{12}$ W/cm$^2$ in the interaction region during the experiment.

\section{\label{sec:level1} Summary and future perspectives}

In conclusion, we have presented a versatile attosecond beamline combining a generation source for generating the attosecond pulses, a photoelectron spectrometer - VMI, and an XUV spectrometer to characterize the spectral amplitude of the generated harmonics. Starting from a CEP-stabilized high-energy source based on Ti:Sa, we have shown our developments to be able to produce few-cycle pulses. We have further shown the ability to generate trains of attosecond pulses and an XUV spectral continuum, which is a basic signature for IAP. Depending on the system under study, the developed beamline can be easily adapted to either trains of attosecond pulses or an IAP. 

Apart from these, the versatility of the developed attosecond beamline was further demonstrated through the field-sampling technique introduced in this work. This method is based on monitoring oscillations in the harmonic yield within the cut-off region of the XUV spectral continuum. To enable phase-sensitive detection via a lock-in technique, a direct readout of the harmonic signal was achieved using a TIA connected to the MCP-P, thereby eliminating the need for additional XUV detection systems. Beyond improving the SNR, this approach allows for higher scan rates, significantly reducing the acquisition time required to record electric-field traces of the Tp. The stability of the system is evidenced by the small $1\sigma$ standard deviation obtained from eight traces measured over approximately 6 hours, reflecting both the spectral stability of the generated few-cycle $5.58$~fs (FWHM) laser pulses and the reproducibility of the field-sampling technique. Furthermore, the high SNR enabled precise control and resolution of the quantum electron trajectories involved in the harmonic generation process. Experimentally, this capability is manifested as clear $2\omega$ oscillations in the harmonic yield. The ability of the field sampling technique to detect Tp with pulse energy as low as a few nanojoules makes this technique an ideal candidate to study the field response in crystals or nanostructures. 

We have also demonstrated photoelectron spectroscopy using VMI spectrometer that will enable future time- and angle-resolved investigation of photoionization processes of different atoms and molecules.

\section*{\label{sec:Acknowldegment} Acknowledgements}
We acknowledge the financial support from the \textit{Deutsche Forschungsgemeinschaft} (DFG, German Research Foundation) Priority Programme 1840, Quantum Dynamics in Tailored Intense Fields (QUTIF), Research Training Group DynCAM (RTG 2717), DFG SA3470/3-1 and DFG SA3470/7-1 and from the European Union’s Horizon Europe research and innovation program under the Marie Skłodowska-Curie grant agreement No 101168628 (project QU-ATTO).

We would like to thank the team at the electrical and mechanical workshops at the Institute of Physics, University of Freiburg, for the respective support in building the setup. 

\section*{\label{sec:References} References}
\bibliography{aipsamp}% Produces the bibliography via BibTeX.

\end{document}